\documentclass[11pt]{article}
\usepackage{blois,epsfig}

\def\be{\begin{equation}}
\def\ee{\end{equation}}
\def\bea{\begin{eqnarray}}
\def\eea{\end{eqnarray}}

\begin{document}
\vspace*{4cm}
\title{RESULTS AND PERSPECTIVES OF THE SOLAR AXION SEARCH WITH THE CAST EXPERIMENT}

\author{ 
\MakeUppercase{
E.FERRER RIBAS$^1$, 
M.~ARIK$^2$,
S.~AUNE$^1$,  
%}\affiliation{\Saclay}
K.~Barth $^3$, 
%}\affiliation{\CERN}
A.~Belov$^4$,
%}\affiliation{\INR}
S.~Borghi$^3$, 
%}\altaffiliation[Present addr.: ]{\Glasgow}\affiliation{\CERN}
H.~Br\"auninger$^5$, 
%}\affiliation{\MPE}
G.~Cantatore$^6$, 
%}\affiliation{\Trieste}
J.~M.~Carmona$^7$, 
%}\affiliation{\Zaragoza}
S.~A.~Cetin$^2$, 
%}\affiliation{\Dogus}
J.~I.~Collar$^8$, 
%}\affiliation{\Chicago}
T.~Dafni $^7$, 
%}\affiliation{\Zaragoza}
M.~Davenport$^3$,
%  }\affiliation{\CERN}
C.~Eleftheriadis$^9$,
%  }\affiliation{\Thessaloniki}
N.~Elias$^3$,
%}\affiliation{\CERN}
C.~Ezer$^2$,
%}\altaffiliation[Present addr.: ]{\Bogazici}\affiliation{\Dogus}
G.~Fanourakis$^{10}$,
% }\affiliation{\Demokritos}
P.~Friedrich$^5$, 
% }\affiliation{\MPE}
J.~Gal\' an$^7$,
%}\altaffiliation[Present addr.: ]{\Saclay}\affiliation{\Zaragoza}
J.~A.~Garc\' ia$^7$,
%}\affiliation{\Zaragoza}
A.~Gardikiotis$^{12}$,
%}\affiliation{\Patras}
J.~G.~Garza$^7$,
E.~N.~Gazis $^{13}$,
%}\affiliation{\Athens}
T.~Geralis$^{10}$,
%}\affiliation{\Demokritos}
I.~Giomataris$^1$,
% }\affiliation{\Saclay}
S.~Gninenko$^4$,
%}\affiliation{\INR}
H.~G\' omez$^7$,
%}\affiliation{\Zaragoza}
E.~Gruber$^{11}$,
%}\affiliation{\Freiburg}
T.~Guth\"orl$^{11}$,
%}\affiliation{\Freiburg}
R.~Hartmann$^{14}$,
%}\altaffiliation[Present addr.: ]{\PNSensor}\affiliation{\MPI}
F.~Haug$^3$,%}\affiliation{\CERN}
M.~D.~Hasinoff$^{15}$, 
%}\affiliation{\Vancouver}
D.~H.~H.~Hoffmann$^{16}$,
%}\affiliation{\Darmstadt}
F.~J.~Iguaz$^1$,
%}\altaffiliation[Present addr.: ]{\Saclay}\affiliation{\Zaragoza}
I.~G.~Irastorza $^7$, 
%}\affiliation{\Zaragoza}
J.~Jacoby$^{17}$, 
%}\affiliation{\Frankfurt}
K.~Jakov\v ci\' c$^{18}$,  
%}\affiliation{\Zagreb}
M.~Karuza$^6$,
%}\affiliation{\Trieste}
K.~K\"onigsmann$^{11}$,
%}\affiliation{\Freiburg}
R.~Kotthaus$^{19}$,
%}\affiliation{\MPP}
M.~Kr\v{c}mar$^{18}$,
 %}\affiliation{\Zagreb}
M.~Kuster$^5$,
%}\altaffiliation[Present addr.: ]{\XFEL}\affiliation{\MPE}\affiliation{\Darmstadt}
B.~Laki\'{c}$^{18}$,
%}\affiliation{\Zagreb}
J.~M.~Laurent$^2$,
%}\affiliation{\CERN}
A.~Liolios$^9$,
%}\affiliation{\loniki}
A.~Ljubi\v{c}i\'{c}$^{18}$,
  %}\affiliation{\Zagreb}
V.~Lozza$^6$, 
%}\affiliation{\Trieste}
G.~Lutz$^{14}$,
%}\altaffiliation[Present addr.: ]{\PNSensor}\affiliation{\MPI}
G.~Luz\'on$^7$,
%}\affiliation{\Zaragoza}
J.~Morales$^7$,
%}\altaffiliation[Deceased.]{}\affiliation{\Zaragoza}
T.~Niinikoski$^2$,
%}\altaffiliation[Present addr.: ]{\ECU}\affiliation{\CERN}
A.~Nordt$^{5,16}$,
%}\altaffiliation[Present addr.: ]{\CERN}\affiliation{\MPE}\affiliation{\Darmstadt}
T.~Papaevangelou$^1$, 
%}\affiliation{\Saclay}
M.~J.~Pivovaroff$^{20}$,
%}\affiliation{\LLNL}
G.~Raffelt$^{19}$,
%}\affiliation{\MPP}
T.~Rashba$^{21}$,
%}\affiliation{\MPIS}
H.~Riege$^{16}$,
%}\affiliation{\Darmstadt}
A.~Rodr\' iguez$^7$,
%}\affiliation{\Zaragoza}
M.~Rosu$^{16}$, 
%}\affiliation{\Darmstadt}
J.~Ruz$^{7,3}$,
%}\affiliation{\Zaragoza}\affiliation{\CERN}
I.~Savvidis$^9$, 
%}\affiliation{\loniki}
P.~S.~Silva$^3$,
%}\affiliation{\CERN}
S.~K.~Solanki$^{21}$, 
%}\affiliation{\MPIS}
L.~Stewart$^3$,
%}\affiliation{\CERN}
A.~Tom\' as$^7$,
%}\affiliation{\Zaragoza}
M.~Tsagri$^{12}$,
%}\altaffiliation[Present addr.: ]{\CERN}\affiliation{\Patras}
K.~van~Bibber$^{20}$, 
%}\altaffiliation[Present addr.: ]{\Naval}\affiliation{\LLNL}
T.~Vafeiadis$^{3,9}$,%}\affiliation{\CERN}\affiliation{\Thessaloniki}\affiliation{\Patras}
J.~Villar$^7$,
%}\affiliation{\Zaragoza}
J.~K.~Vogel$^{20}$,
%}\altaffiliation[Present addr.: ]{\LLNL}\affiliation{\Freiburg}\affiliation{\LLNL}
S.~C.~Yildiz$^2$,
%}\altaffiliation[Present addr.: ]{\Bogazici}\affiliation{\Dogus}
K.~Zioutas$^{12}$   (CAST Collaboration) %}\affiliation{\CERN}\affiliation{\Patras}
}
}

\address{
1.IRFU, CEA, Centre de Saclay, 91191 Gif sur Yvette, France\\
2.Dogus University, Istanbul, Turkey\\
3.European Organization for Nuclear Research (CERN), Gen\`eve, Switzerland \\
4.Institute for Nuclear Research (INR), Russian Academy of Sciences, Moscow, Russia\\
5.Max-Planck-Institut f\"{u}r Extraterrestrische Physik, Garching, Germany\\
6.Instituto Nazionale di Fisica Nucleare (INFN), Sezione di Trieste and Universit\`a di Trieste, Trieste, Italy\\
7.Instituto de F\'{\i}sica Nuclear y Altas Energ\'{\i}as, Universidad de Zaragoza, Zaragoza, Spain \\
8.Enrico Fermi Institute and KICP, University of Chicago, Chicago, IL, USA\\
9.Aristotle University of Thessaloniki, Thessaloniki, Greece\\
10.National Center for Scientific Research ``Demokritos'', Athens, Greece\\
11.Albert-Ludwigs-Universit\"{a}t Freiburg, Freiburg, Germany\\
12.Physics Department, University of Patras, Patras, Greece\\
13.National Technical University of Athens, Athens, Greece\\
14.MPI Halbleiterlabor, M\"{u}nchen, Germany\\
15.Department of Physics and Astronomy, University of British Columbia, Vancouver, Canada \\
16.Technische Universit\"{a}t Darmstadt, IKP, Darmstadt, Germany\\
17.Johann Wolfgang Goethe-Universit\"at, Institut f\"ur Angewandte Physik, Frankfurt am Main, Germany\\
18.Rudjer Bo\v{s}kovi\'{c} Institute, Zagreb, Croatia\\
19.Max-Planck-Institut f\"{u}r Physik (Werner-Heisenberg-Institut), M\"unchen, Germany\\
20.Lawrence Livermore National Laboratory, Livermore, CA, USA\\
21.Max-Planck-Institut f\"{u}r Sonnensystemforschung, Katlenburg-Lindau, Germany\\
}

\maketitle\abstracts{
The status of the solar axion search with the CERN Axion Solar Telescope (CAST) will be presented. Recent results obtained by the use of $^3$He as a buffer gas has allowed us to extend our sensitivity to higher axion masses than our previous measurements with $^4$He. With about 1 h of data taking at each of 252 different pressure settings we have scanned the axion mass range 0.39 eV$ \le m_{a} \le $ 0.64 eV. From the absence of an excess of x rays when the magnet was pointing to the Sun we set a typical upper limit on the axion-photon coupling of g$_{a\gamma} \le 2.3\times 10^{-10}$  GeV$^{-1}$ at 95\% C.L., the exact value depending on the pressure setting. CAST published results represent the best experimental limit on the photon couplings to axions and other similar exotic particles dubbed WISPs (Weakly Interacting Slim Particles) in the considered mass range and for the first time the limit enters the region favored by QCD axion models. Preliminary sensitivities for axion masses up to 1.16 eV will also be shown reaching  mean upper limits  on the axion-photon coupling of g$_{a\gamma} \le 3.5\times 10^{-10}$  GeV$^{-1}$ at 95\% C.L. Expected sensibilities for the extension of the CAST program up to 2014 will be presented.  Moreover long term options for a new helioscope experiment will be evoked.
}

\section{The CAST sensitivity}
The CAST (Cern Axion Solar Telescope)\cite{Zio05,And07} experiment is using a
decommissioned LHC dipole magnet to convert solar axions into
detectable x-ray photons. Axions are light pseudoscalar particles
that arise in the context of the Peccei-Quinn\cite{Peccei}
solution to the strong CP problem and can be Dark Matter
candidates\cite{Sikivie}. Stars could produce axions via the
Primakoff conversion of the plasma photons. The CAST experiment
is pointing at our closest star, the Sun, aiming to detect solar axions. The
detection principle is based on the coupling of an incoming axion
to a virtual photon provided by the transverse field of an intense
dipole magnet, being transformed into a real, detectable photon
that carries the energy and the momentum of the original axion.
The axion to photon conversion probability is proportional to the
square of the transverse field of the magnet and to the active
length of the magnet. Using an LHC magnet ($9\;$T and $9.26\;$m long)
improves the conversion probability by a factor 100 compared to previous
experiments. 

Thanks to its rotating platform CAST can point and track the Sun during 3 h while sunrise and sunset time. The rest of the day is devoted to measurements of background. Two different X-ray detectors are used presently. Three Micromegas detectors\cite{castmm,mmcast1,mmcast2} cover the two bores looking for sunset axions and one of the bores of the sunrise side.  A CCD\cite{castccd} is covering the forth emplacement in the sunrise side. One of the originalities of the experiment is the use of an X-ray focusing mirror system, designed and built as a spare for the X-ray astronomy mission ABRIXAS. It provides a focusing of the X-rays coming out of the magnet down to a spot of a few mm$^2$ on the CCD, further increasing the signal-to-noise ratio and therefore the sensitivity of the experiment.

\begin{figure}
\centering%
\begin{tabular}{cc}
\includegraphics[width=0.45\textwidth]{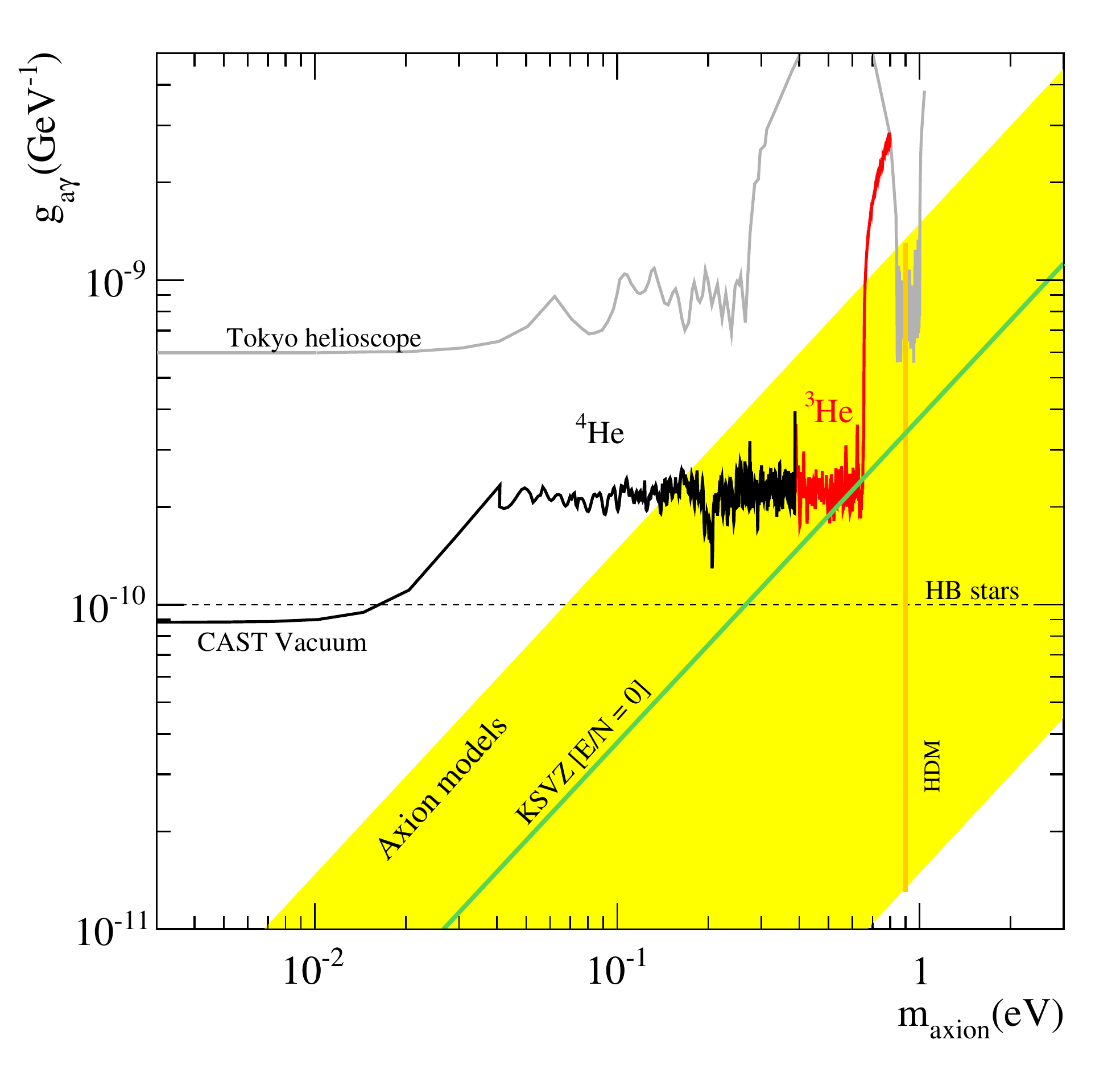}
\includegraphics[width=0.57\textwidth]{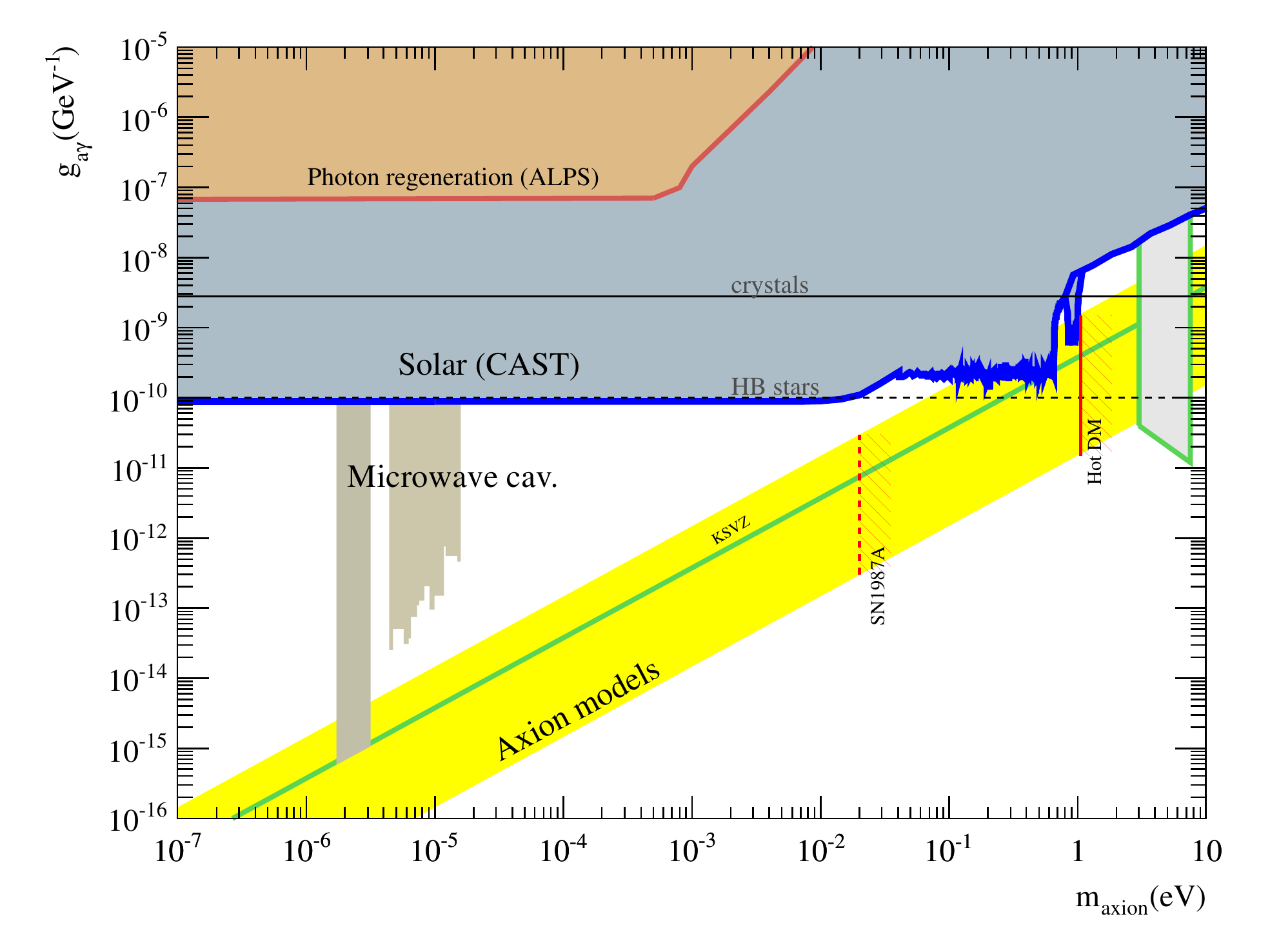}
\end{tabular}
\caption{\textit{Left: Latest CAST limit on g$_{a\gamma}$ as a function of m$_a$ obtained with the first part the $^3$He data reaching a sensitivity up to 0.64~eV. Right: CAST limits compared to other measurements and bound from theoretical, astrophysical as well as cosmologically derived upper (HDM). The blue line shows the 95\% exclusion limit from published CAST (vacuum phase and $^{4}$He data). The red line shows the 95\% exclusion limit obtained from the analysis of the 2008 $^{3}$He data which has been published in PRL. The Tokyo helioscope limits are also shown. The yellow band indicates the favoured theoretical region for axion models. }}
\label{fig:todaylimits}
\end{figure}

The CAST experiment has been taking data since 2003 providing the most restrictive experimental limits on the axion-photon coupling for a broad range of axion masses\cite{cast1,cast2,cast3,cast4}. In 2003 and 2004 the experiment operated with vacuum inside the magnet (CAST phase I) and set the best experimental limit on the axion-photon coupling constant in the range of axion masses up to 0.02 eV. Beyond this mass the sensitivity is degraded due to coherence loss. In order to restore coherence, the magnet can be filled with a buffer gas providing an effective mass to the photon. By changing the pressure of the buffer gas in steps, one can scan an entire range of axion mass values.  The CAST experiment started this gas program entering its phase II at the end of 2005. From 2005 to 2007, the magnet bore was filled with $^4$He gas extending the sensitivity to masses up to 0.4 eV. From March 2008 onwards the magnet bore was been filled with $^3$He. With the end of the 2011 data taking in July, the CAST experiment has covered axion masses up to 1.18 eV surpassing the initial goal of the phase II which was to reach 1.16 eV. The results of the first part of the $^3$He data, with a sensitivity up to 0.64 eV, have been published in \cite{cast4} and are given in figure~\ref{fig:todaylimits} (left). In figure~\ref{fig:todaylimits} (right) a more general plot shows the current experimental panorama. The three main front lines of direct detection experiments are highlited: laser-based laboratory techniques, helioscope (solar ALPs (Axion Like Particles) and axions),and microwave cavities (dark matter axions). The blue line corresponds to the current helioscope limits, dominated by CAST for practically all axion masses but for the m$_a \sim $0.85-1 eV exclusion line from the last Tokyo helioscope results\cite{tokyo}. Also shown are the constraints from horizontal branch (HB) stars, supernova SN1987A, and hot dark matter (HDM). The yellow "axion band" represents the range of realistic models. 
The analysis of the data covering masses up to 1.18 eV is in progress and preliminary results were shown in the presentation. 

The collaboration has performed by-product analysis of the data taken, to look for other
axion scenario to which CAST would also be sensitive. The TPC phase I data has been reanalysed in order look for 14 keV axions coming from M1 transitions\cite{14keV}. In addition, data taken with a calorimeter during the phase I, 
were used to search for high energy (MeV) lines from high energy axion conversion\cite{Kresso}. Moreover a few days of data were taken with a visible detector coupled to one end of the CAST magnet\cite{visible}, in search for axions with energy in the "visible" range. A permanent setup has been installed in the experiment in order to take data without interfering with the standard program of CAST.

\section{Short and long terms prospects: CAST up to 2014 and IAXO}
The CAST Collaboration has decided to extend its program up to 2014 in order to profit from the fact that the Micromegas detectors presently running have a  factor 20 better background level than at the beginning of the experiment in 2002. This fact allows us to consider revisiting some of the past data taking configurations with enhanced sensitivity to standard Peccei-Quinn axion models, the main objective of CAST. In 2012 the run is devoted to revisit phase II $^4$He to enhance the sensitivity in the region of around 0.4 eV. This gain
in sensitivity is expected from our current detectors that exhibit much better performances and at the same time
the stepping strategy that will focus on a restricted mass range but with increased statistics per density step. This
will allow to improve our current sensitivity obtained with $^4$He run and to cross the benchmark KSVZ axion models.
In parallel we are carrying out an ambitious R \& D program to be ready for 2013 and 2014 where we would like to
take data with vacuum in the magnet bores as in CAST phase I. To attain a significant gain in sensitivity
we need detectors reaching background levels of the order of $10^{-7}$counts\,keV$^{-1}$\,cm$^{-2}$\,s$^{-1}$ and new X-ray optics. We
have a roadmap for Micromegas detectors to reach these background levels and a project of developing one new X-ray telescope to be installed in the Sunrise Micromegas line.

In order to reach further sensitivities a new axion helioscope, IAXO (International AXion Observatory)\cite{NGAH}, is being designed. The key ingredients are: a large and strong magnet, focusing optics and low background detectors. A complete feasibility study is currently in progress to optimise the magnet design that satisfies the requirements of IAXO. The toroidal design  seems to be the simplest and cheapest candidate to improve significantly the sensitivity. Concerning the focussing optics, plastic-substrate techniques are the most promising.
As regards to detectors, the IAXO proposal is based on the performance of the present Micromegas detectors with different improvements related to the optimisation of the shielding, the replacement of the present electronics by TPC-like electronics  and the selection of non radioactive materials.

The aim of such experiment is to improve by a factor 10-30 the CAST sensitivity. Except for the axion dark matter searches in a narrow axion mass range, this experiment will be the most sensitive axion search ever, reaching or surpassing the stringent bounds from SN1987A and possibly testing the axion interpretation of anomalous white-dwarf cooling that predicts $\rm{m_a}$ of a few meV. 

%\section{Conclusion}

\end{document}